\documentstyle[12pt,epsfig]{article}
\begin{document}

\vspace{0.2cm}

\begin{center}
{\large\bf Implications of Leptonic Unitarity Violation at Neutrino
Telescopes}
\end{center}

\vspace{0.5cm}
\begin{center}
{\bf Zhi-zhong Xing} \footnote{E-mail: xingzz@ihep.ac.cn} ~ and ~
{\bf Shun Zhou} \footnote{E-mail: zhoush@ihep.ac.cn} \\
\vspace{0.2cm} {\small\sl Institute of High Energy Physics, Chinese
Academy of Sciences, Beijing 100049, China}
\end{center}

\vspace{3.5cm}

\begin{abstract}
A measurement of the ultrahigh-energy (UHE) cosmic neutrinos at a
km$^3$-size neutrino telescope will open a new window to constrain
the $3\times 3$ neutrino mixing matrix $V$ and probe possible new
physics. We point out that it is in principle possible to examine
the non-unitarity of $V$, which is naturally expected in a class
of seesaw models with one or more TeV-scale Majorana neutrinos, by
using neutrino telescopes. Considering the UHE neutrinos produced
from the decays of charged pions arising from $pp$ and (or)
$p\gamma$ collisions at a distant astrophysical source, we show
that their flavor ratios at a terrestrial neutrino telescope may
deviate from the democratic flavor distribution $\phi^{\rm T}_e :
\phi^{\rm T}_\mu : \phi^{\rm T}_\tau = 1 : 1 : 1$ due to the
seesaw-induced unitarity violation of $V$. Its effect can be as
large as several percent and can serve for an illustration of how
sensitive a neutrino telescope should be to this kind of new
physics.
\end{abstract}

\newpage

\section{Introduction}

The solar \cite{SNO}, atmospheric \cite{SK}, reactor \cite{KM} and
accelerator \cite{K2K} neutrino experiments have provided us with
very convincing evidence that neutrinos are massive and lepton
flavors are mixed. In the basis where the flavor eigenstates of
charged leptons coincide with their mass eigenstates, the phenomenon
of neutrino mixing can simply be described by a $3\times 3$ unitary
matrix $V$ which links the neutrino flavor eigenstates $(\nu^{}_e,
\nu^{}_\mu, \nu^{}_\tau)$ to the neutrino mass eigenstates
$(\nu^{}_1, \nu^{}_2, \nu^{}_3)$:
\begin{equation}
\left ( \matrix{\nu^{}_e \cr \nu^{}_\mu \cr \nu^{}_\tau \cr} \right
) \; =\; \left ( \matrix{ V^{}_{e1} & V^{}_{e2} & V^{}_{e3} \cr
V^{}_{\mu 1} & V^{}_{\mu 2} & V^{}_{\mu 3} \cr V^{}_{\tau 1} &
V^{}_{\tau 2} & V^{}_{\tau 3} \cr} \right ) \left ( \matrix{
\nu^{}_1 \cr \nu^{}_2 \cr \nu^{}_3 \cr} \right ) \; .
\end{equation}
A full parametrization of $V$ requires 3 rotation angles
$(\theta^{}_{12}, \theta^{}_{13}, \theta^{}_{23})$ and 3 phase
angles $(\delta, \rho, \sigma)$ \cite{PDG}:
\begin{eqnarray}
V = \left(\matrix{c^{}_{13} c^{}_{12} & c^{}_{13} s^{}_{12} & ~
s^{}_{13}e^{-i\delta} \cr -s^{}_{12} c^{}_{23} - c^{}_{12}
s^{}_{23} s^{}_{13}e^{i\delta} & + c^{}_{12} c^{}_{23} - s^{}_{12}
s^{}_{23} s^{}_{13}e^{i\delta} & c^{}_{13} s^{}_{23} \cr +
s^{}_{12} s^{}_{23} - c^{}_{12} c^{}_{23} s^{}_{13}e^{i\delta} &
-c^{}_{12} s^{}_{23} - s^{}_{12} c^{}_{23} s^{}_{13}e^{i\delta} &
c^{}_{13} c^{}_{23}}\right) P^{}_{\rm M} \; ,
\end{eqnarray}
where $s^{}_{ij} \equiv \sin \theta^{}_{ij}, c^{}_{ij} \equiv \cos
\theta^{}_{ij}$ (for $ij=12, 13, 23$), and $P^{}_{\rm M} = {\rm
Diag}\{1, e^{i\rho}, e^{i\sigma}\}$ is the Majorana phase matrix
irrelevant to neutrino oscillations. A global analysis of current
experimental data \cite{Vissani} points to $\theta^{}_{13} \approx
0$ and $\theta^{}_{23} \approx \pi/4$, a noteworthy result which
has motivated a number of authors to consider the $\mu$-$\tau$
permutation symmetry and its breaking mechanism for model building
\cite{Symmetry}.

Now that neutrinos can oscillate from one flavor to another, it
will be extremely interesting to detect the oscillatory phenomena
of ultrahigh-energy (UHE) cosmic neutrinos produced from distant
astrophysical sources. IceCube \cite{IceCube}, a ${\rm
km}^3$-volume under-ice neutrino telescope, is now under
construction at the South Pole and aims to observe the UHE
neutrino oscillations. Together with the under-water neutrino
telescopes in the Mediterranean Sea (ANTARES \cite{ANTARES},
NESTOR \cite{NESTOR} and NEMO \cite{NEMO}), IceCube has the
potential to shed light on the acceleration mechanism of UHE
cosmic rays and to probe the intrinsic properties of cosmic
neutrinos. An immediate consequence of neutrino oscillations is
that the flavor composition of cosmic neutrinos to be observed at
the telescopes must be different from that at the sources
\cite{Pakvasa}. By measuring the cosmic neutrino flavor
distribution, one can determine or constrain the mixing angles
$(\theta^{}_{12}, \theta^{}_{13}, \theta^{}_{23})$ and the Dirac
CP-violating phase $(\delta$). A lot of attention has recently
been paid to this intriguing possibility
\cite{Serpico}---\cite{decay}.

We aim to investigate the oscillation of cosmic neutrinos produced
from the decays of charged pions arising from energetic $pp$ and
(or) $p\gamma$ collisions at a distant astrophysical source (e.g.,
active galactic nuclei or AGN). For such a most probable UHE
neutrino source, its flavor composition is
\begin{eqnarray}
\phi^{}_{e} : \phi^{}_\mu : \phi^{}_\tau = 1 : 2 : 0 \; ,
\end{eqnarray}
where $\phi^{}_\alpha \equiv \phi^{}_{\nu^{}_\alpha} +
\phi^{}_{\bar{\nu}^{}_\alpha}$ (for $\alpha = e, \mu, \tau$) denotes
the $\alpha$-neutrino flux at the source. As the distances between
the astrophysical sources and the terrestrial detectors are much
longer than the typical length of solar or atmospheric neutrino
oscillations, one may average the UHE cosmic neutrino oscillation
probabilities and arrive at
\begin{eqnarray}
P^{}_{\alpha \beta} \equiv P(\nu^{}_\alpha \to \nu^{}_\beta) =
\sum^3_{i=1} |V^{}_{\alpha i}|^2 |V^{}_{\beta i}|^2 \; .
\end{eqnarray}
This result is also valid for the anti-neutrino oscillations;
namely, $\bar{P}_{\alpha \beta} \equiv P(\bar{\nu}^{}_\alpha \to
\bar{\nu}^{}_\beta) = P^{}_{\alpha \beta}$ for $\alpha, \beta = e,
\mu$ and $\tau$. Therefore, the neutrino fluxes at the detector
can be calculated from
\begin{eqnarray}
\phi^{\rm T}_\alpha = \sum_\beta P^{}_{\alpha \beta} \phi^{}_\beta
\; .
\end{eqnarray}
Given Eq. (3) together with the condition $|V^{}_{\mu i}| =
|V^{}_{\tau i}|$ (for $i=1,2,3$) \cite{XZ08}, it is easy to show
that the flavor distribution of UHE cosmic neutrinos has a
democratic pattern at neutrino telescopes:
\begin{eqnarray}
\phi^{\rm T}_{e} : \phi^{\rm T}_\mu : \phi^{\rm T}_\tau = 1 : 1 :
1 \; .
\end{eqnarray}
Note that $|V^{}_{\mu i}| = |V^{}_{\tau i}|$ implies either
$\theta^{}_{13} = 0$ and $\theta^{}_{23} = \pi/4$ (CP invariance)
or $\delta=\pm \pi/2$ and $\theta^{}_{23} = \pi/4$ (CP violation)
in the standard parametrization of $V$ as shown in Eq. (2). These
two sets of interesting conditions can be realized from the
so-called tri-bimaximal \cite{TB} and tetra-maximal \cite{TT}
neutrino mixing scenarios, respectively.

One has to bear in mind that $\phi^{\rm T}_e : \phi^{\rm T}_\mu :
\phi^{\rm T}_\tau = 1 : 1 : 1$ depends on two idealized
hypotheses: the astrophysical source of UHE neutrinos satisfies
$\phi^{}_{e} : \phi^{}_\mu : \phi^{}_\tau = 1 : 2 : 0$ and the
$3\times 3$ neutrino mixing matrix $V$ satisfies $|V^{}_{\mu i}| =
|V^{}_{\tau i}|$. Previous works have extensively analyzed
possible deviations from the democratic flavor distribution of UHE
cosmic neutrinos at neutrino telescopes by taking account of the
energy dependence, uncertainties in the neutrino mixing angles,
contaminations to the canonical production of $\nu^{}_e$'s
($\bar{\nu}^{}_e$'s) and $\nu^{}_\mu$'s ($\bar{\nu}^{}_\mu$'s)
from $\pi^\pm$'s, and different sources of UHE cosmic neutrinos
\cite{Serpico}---\cite{decay}.

We shall concentrate on the standard pion-decay source of UHE
neutrinos, whose flavor composition has been given in Eq. (3), to
explore the effects of non-unitarity of $V$ on the flavor
distribution of such cosmic neutrinos at a terrestrial neutrino
telescope. This investigation is new and makes sense, because $V$
is naturally expected to be non-unitary in a class of seesaw
models with one or more TeV-scale right-handed Majorana neutrinos.
We find that the democratic flavor distribution in Eq. (6) can be
broken at the percent level as a consequence of the unitarity
violation of $V$. Although such a small effect is hard to be
observed in any realistic experiments in the foreseeable future,
it {\it does} illustrate how sensitive a neutrino telescope should
be to this kind of new physics.

\section{Unitarity Violation at Neutrino Telescopes}

If the tiny masses of three known neutrinos ($\nu^{}_1, \nu^{}_2,
\nu^{}_3$) are attributed to the popular seesaw mechanism (either
type-I \cite{SS1} or type-II \cite{SS2}), in which there exist a few
heavy (right-handed) Majorana neutrinos $N^{}_i$, then the $3\times
3$ neutrino mixing matrix $V$ must be non-unitary. The effect of
unitarity violation of $V$ depends on the mass scale of $N^{}_i$,
and it can be of ${\cal O}(10^{-2})$ if $N^{}_i$ are at the TeV
scale \cite{XZ} --- an energy frontier to be explored by the LHC.
Indeed, a global analysis of current neutrino oscillation data and
precision electroweak data yields some stringent constraints on the
non-unitarity of $V$, but its effect is allowed to be of ${\cal
O}(10^{-2})$ \cite{Antusch} and may have some novel implications on
neutrino oscillations \cite{Xing08}---\cite{Zralek}.

In the presence of small unitarity violation, we write the
neutrino mixing matrix as $V = A V^{}_0$, where $V^{}_0$ is a
unitary matrix containing 3 rotation angles ($\theta^{}_{12},
\theta^{}_{13}, \theta^{}_{23}$) and 3 phase angles like that
given in Eq. (2), and $A$ is a quasi-identity matrix which can in
general be parametrized in terms of 9 rotation angles
$\theta^{}_{ij}$ and 9 phase angles $\delta^{}_{ij}$ (for $i
=1,2,3$ and $j=4,5,6$) \cite{Xing08}. For simplicity, here we
adopt the expression of $A$ shown in Eq. (11) of Ref.
\cite{Xing08} and take $V^{}_0$ to be the well-known tri-bimaximal
mixing pattern \cite{TB} without any CP-violating phases. Then we
obtain the non-unitary neutrino mixing matrix $V = AV^{}_0$ as
follows:
\begin{eqnarray}
V = \left(\matrix{\frac{2}{\sqrt{6}} \left(1 - W^{}_1\right) &
\frac{1}{\sqrt{3}} \left(1 - W^{}_1\right) & 0 \cr
-\frac{1}{\sqrt{6}} \left(1 - W^{}_2 + 2X\right) &
\frac{1}{\sqrt{3}} \left(1 - W^{}_2 - X\right) &
\frac{1}{\sqrt{2}} \left(1 - W^{}_2\right) \cr \frac{1}{\sqrt{6}}
\left(1 - W^{}_3 - 2Y + Z\right) & -\frac{1}{\sqrt{3}} \left(1 -
W^{}_3 + Y + Z\right) & \frac{1}{\sqrt{2}} \left(1 - W^{}_3 -
Z\right)}\right) \; ,
\end{eqnarray}
where
\begin{eqnarray}
W^{}_i = \frac{1}{2} \left(s^2_{i4} + s^2_{i5} + s^2_{i6}\right)
\; ,
\end{eqnarray}
for $i = 1, 2, 3$; and
\begin{eqnarray}
X & = & \hat{s}^{}_{14} \hat{s}^*_{24} + \hat{s}^{}_{15}
\hat{s}^*_{25} + \hat{s}^{}_{16} \hat{s}^*_{26} \; , \nonumber \\
Y & = & \hat{s}^{}_{14} \hat{s}^*_{34} + \hat{s}^{}_{15}
\hat{s}^*_{35} + \hat{s}^{}_{16} \hat{s}^*_{36} \; , \nonumber \\
Z & = & \hat{s}^{}_{24} \hat{s}^*_{34} + \hat{s}^{}_{25}
\hat{s}^*_{35} + \hat{s}^{}_{26} \hat{s}^*_{36} \; .
\end{eqnarray}
Here $s^{}_{ij} \equiv \sin\theta^{}_{ij}$ and $\hat{s}^{}_{ij}
\equiv e^{i\delta^{}_{ij}} s^{}_{ij}$ have been defined, and
higher-order terms of $s^{}_{ij}$ have been neglected. The mixing
angles in $\theta^{}_{ij}$ can at most be of ${\cal O}(0.1)$, but
the CP-violating phases $\delta^{}_{ij}$ are entirely unrestricted.
If both $\theta^{}_{ij}$ and $\delta^{}_{ij}$ are switched off, the
tri-bimaximal neutrino mixing pattern will be reproduced from Eq.
(7). With the help of Eqs. (4) and (7), we arrive at
\begin{eqnarray}
P^{}_{ee} &=& \frac{5}{9} - \frac{20}{9} W^{}_1 \; ,\nonumber \\
P^{}_{e \mu} &=& \frac{2}{9} - \frac{4}{9} \left(W^{}_1 +
W^{}_2\right) + \frac{2}{9} {\rm Re} X \; , \nonumber \\
P^{}_{e \tau} &=& \frac{2}{9} - \frac{4}{9} \left(W^{}_1 +
W^{}_3\right) - \frac{2}{9} \left({\rm Re}Y - 2{\rm Re}Z\right) \;
, \nonumber \\
P^{}_{\mu \mu} &=& \frac{7}{18} - \frac{14}{9}W^{}_2 -
\frac{2}{9}{\rm Re}X \; , \nonumber \\
P^{}_{\mu \tau} &=& \frac{7}{18} - \frac{7}{9} \left(W^{}_2 +
W^{}_3\right) - \frac{1}{9} \left({\rm Re}X - {\rm Re}Y + 2{\rm
Re}Z\right) \; , \nonumber \\
P^{}_{\tau \tau} &=& \frac{7}{18} - \frac{14}{9}W^{}_3 + \frac{2}{9}
\left({\rm Re}Y - 2{\rm Re}Z\right) \; .
\end{eqnarray}
For the canonical astrophysical source of UHE neutrinos under
consideration, we definitely have $\{\phi^{}_e, \phi^{}_{\mu},
\phi^{}_{\tau}\} = \{1/3, 2/3, 0\} \phi^{}_0$, where $\phi^{}_0$
denotes the total initial flux. It is then easy to get the flavor
distribution at a terrestrial neutrino telescope:
\begin{eqnarray}
\phi^{\rm T}_e &\equiv& \frac{\phi^{}_0}{3} \left[1 -
\frac{4}{9}\left(7W^{}_1 + 2W^{}_2\right) + \frac{4}{9}{\rm
Re}X\right] \; , \nonumber \\
\phi^{\rm T}_\mu &\equiv& \frac{\phi^{}_0}{3} \left[1 - \frac{4}{9}
\left(W^{}_1 + 8W^{}_2\right) - \frac{2}{9}{\rm Re}X\right] \; ,
\nonumber \\
\phi^{\rm T}_\tau &\equiv& \frac{\phi^{}_0}{3} \left[1 - \frac{2}{9}
\left(2W{}_1 + 7W^{}_2 + 9W^{}_3\right) - \frac{2}{9}{\rm Re}X
\right] \; .
\end{eqnarray}
The democratic flavor distribution of $\phi^{\rm T}_\alpha$ (for
$\alpha =e, \mu, \tau$) is clearly broken. Because of the
non-unitarity of $V$, the total flux of UHE cosmic neutrinos at
the telescope is not equal to that at the source:
\begin{eqnarray}
\sum_\alpha \phi^{\rm T}_\alpha = \phi^{}_0 \left[1 - \frac{2}{3}
\left(2W^{}_1 + 3W^{}_2 + W^{}_3\right)\right] \; .
\end{eqnarray}
This sum is apparently smaller than $\phi^{}_0$, and it
approximately amounts to $0.96\phi^{}_0$ if $W{}_i \sim 0.01$ (for
$i = 1, 2, 3$). Some comments are in order.

(1) Note that ${\rm Re}X$ receives the most stringent constraint
from current experimental data, $|X| < 7.0\times 10^{-5}$
\cite{Antusch}. Hence the dominant effects of unitarity violation
on $\phi^{\rm T}_\alpha$ come from $W^{}_{i}$. The breaking of
$\phi^{\rm T}_{e} : \phi^{\rm T}_\mu : \phi^{\rm T}_\tau = 1 : 1 :
1$ can be as large as several percent. Although the strength of
unitarity violation is very small and certainly difficult to be
observed in realistic experiments, it {\it does} illustrate how
sensitive a neutrino telescope should be to this kind of new
physics.

(2) Note also that the oscillation probabilities of UHE cosmic
neutrinos are actually given by $\hat{P}^{}_{\alpha \beta} \equiv
P^{}_{\alpha \beta}/\left[(VV^\dagger) ^{}_{\alpha \alpha}
(VV^\dagger)^{}_{\beta \beta}\right]$ (for $\alpha, \beta = e,
\mu, \tau$) in the non-unitary case, where the production of
$\nu^{}_\alpha$ and the detection of $\nu^{}_\beta$ are both
governed by the charged-current interactions \cite{Antusch}. Given
the canonical source of UHE neutrinos, $\nu^{}_e$'s are generated
from the decay of muons, and thus the charged-current interaction
involves two lepton flavors (i.e., $e$ and $\mu$). But
$\nu^{}_\mu$'s can be produced from two channels: one is the decay
of charged pions and the other is the decay of muons. The former
involves only one lepton flavor (i.e., $\mu$). Hence one should
take care of the normalization factors when doing specific
calculations of the cosmic neutrino fluxes for a specific
neutrino-telescope experiment. For the simple pattern of $V$ taken
above, the normalization factors can be explicitly written as
\begin{eqnarray}
VV^\dagger = {\bf 1} - \left(\matrix{2W^{}_1 & X^* & Y^* \cr X &
2W^{}_2 & Z^*  \cr Y & Z & 2W^{}_3}\right) \; .
\end{eqnarray}

(3) The unitarity violation of $V$ under discussion is ascribed to
the existence of heavy Majorana neutrinos in seesaw models and
usually referred to as the minimal unitarity violation
\cite{Antusch}. In contrast, the existence of one or more light
sterile neutrinos and their mixing with three active neutrinos may
also violate the unitarity of $V$. Using $S^{}_{\alpha j}$ to
denote the matrix elements of active-sterile neutrino mixing, we
can express the averaged probabilities of UHE cosmic neutrino
oscillations as
\begin{equation}
P^{}_{\alpha \beta} \equiv P(\nu^{}_\alpha \rightarrow
\nu^{}_\beta) = \sum^3_{i = 1} |V^{}_{\alpha i}|^2 |V^{}_{\beta
i}|^2 + \sum^{n}_{j = 1} |S^{}_{\alpha j}|^2 |S^{}_{\beta j}|^2 \;
,
\end{equation}
where $\alpha$ and $\beta$ run over $e$, $\mu$ and $\tau$, and
\begin{equation}
\sum^3_{i = 1} |V^{}_{\alpha i}|^2 + \sum^{n}_{j = 1}
|S^{}_{\alpha j}|^2 = 1 \; , ~~~ ({\rm for} ~ n = 1, 2, \cdots) \;
\end{equation}
holds. Eq. (15) shows the apparent unitarity violation of $V$
induced by light sterile neutrinos. Two observations have been
achieved in Ref. \cite{Sterile}: (a) for small active-sterile
mixing (i.e., $|S^{}_{\alpha j}| \ll 1$), the effect of
non-unitarity of $V$ at neutrino telescopes is very small and
quite similar to that obtained in Eq. (10); (b) for large
hitherto-unconstrained mixing between active and sterile neutrino
species (i.e., $|S^{}_{\alpha j}| \leq 1$), the existence of light
sterile neutrinos might significantly modify the democratic flavor
distribution of UHE cosmic neutrinos at neutrino telescopes. At
present, however, we have to admit that there is no strong
experimental or theoretical motivation to introduce light sterile
neutrinos into the standard model.

For illustration, we simply assume that there is only one heavy
Majorana neutrino, which can be accommodated in the minimal
type-II seesaw model \cite{Mtype2}. In this case, we are left with
three mixing angles $(\theta^{}_{14}, \theta^{}_{24},
\theta^{}_{34})$ and three CP-violating phases $(\delta^{}_{14},
\delta^{}_{24}, \delta^{}_{34})$ characterizing the unitarity
violation of $V$. As done in Ref. \cite{Xing06}, three working
observables at a neutrino telescope can be defined:
\begin{eqnarray}
R^{}_e &\equiv& \frac{\phi^{\rm T}_e}{\phi^{\rm T}_{\mu} + \phi^{\rm
T}_{\tau}} \; , \nonumber \\
R^{}_\mu &\equiv& \frac{\phi^{\rm T}_\mu}{\phi^{\rm T}_e + \phi^{\rm
T}_{\tau}} \; , \nonumber \\
R^{}_\tau &\equiv& \frac{\phi^{\rm T}_\tau}{\phi^{\rm T}_e +
\phi^{\rm T}_{\mu}} \; .
\end{eqnarray}
In the unitarity limit where $V$ takes the tri-bimaximal mixing
pattern, one can easily obtain $R^{}_e = R^{}_\mu = R^{}_\tau =
1/2$, a result which is equivalent to the democratic flavor
distribution. With the help of Eqs. (8), (9) and (11), we are able
to evaluate the above flux ratios in the presence of unitarity
violation:
\begin{eqnarray}
R^{}_e &\approx& \frac{1}{2} - \frac{1}{36} \left[24s^2_{14} -
15s^2_{24} - 9s^2_{34} - 12 s^{}_{14} s^{}_{24} \cos \varrho \right] \; , \nonumber \\
R^{}_\mu &\approx& \frac{1}{2} + \frac{1}{36} \left[12s^2_{14} -
21s^2_{24} + 9s^2_{34} - 6s^{}_{14} s^{}_{24} \cos \varrho \right] \; , \nonumber \\
R^{}_\tau &\approx& \frac{1}{2} + \frac{1}{36} \left[12s^2_{14} +
6s^2_{24} - 18s^2_{34} - 6s^{}_{14} s^{}_{24} \cos \varrho \right]
\; ,
\end{eqnarray}
where $\varrho \equiv \delta^{}_{14} - \delta^{}_{24}$ and the
higher-order terms of $s^{}_{ij}$ (for $ij = 14, 24, 34$) have
been neglected. Taking into account the experimental constraints
\cite{Antusch}, we have numerically calculated the allowed regions
of these working observables in Figure 1, where the phase angle
$\varrho$ varies freely in the range $\varrho \in [0, 2\pi]$. Two
comments are in order:
\begin{itemize}
\item The deviation of $R^{}_\alpha$ (for $\alpha = e, \mu, \tau$)
from its value in the unitarity limit (i.e., $R^{}_\alpha = 1/2$) is
at most at the $0.1\%$ level. There are two obvious reasons for this
result: (a) there exist significant cancellations among the
contributions of three mixing angles to the flavor ratios; (b) the
mixing angles $s^{}_{14}$ and $s^{}_{24}$ are strictly constrained
by $|X| = s^{}_{14} s^{}_{24} < 7.0 \times 10^{-5}$.

\item In more general cases with two or three heavy Majorana
neutrinos, the above constraint can be loosened. Taking two
TeV-scale Majorana neutrinos for example, we can obtain $s^{}_{ij}
\sim 0.1$ (for $i = 1, 2, 3$ and $j=4, 5$) when the destructive
interference between $\hat{s}^{}_{14} \hat{s}^*_{24}$ and
$\hat{s}^{}_{15} \hat{s}^*_{25}$ terms takes place in $X$ (see Eq.
(9) and switch off the contribution of $\hat{s}^{}_{16}
\hat{s}^*_{26}$ to $X$).
\end{itemize}
While a neutrino telescope is expected to identify different
flavors of UHE cosmic neutrinos, it is also expected to measure
the total flux as precisely as possible. A notable feature of
unitarity violation of $V$ is that the total flux at the detector
is not equal to that at the source, and such a discrepancy may be
as large as several percent shown in Eq. (12).

\section{Comments on cosmic neutrino decays}

So far we have assumed cosmic neutrinos to be stable particles and
studied their flavor distribution at neutrino telescopes. Now let us
make some comments on cosmic neutrino decays and their possible
signatures at neutrino telescopes. It is actually {\it not}
unnatural to speculate that massive neutrinos are unstable and can
decay into lighter neutrinos and other massless particles. If
neutrino masses arise from spontaneous breaking of the global
$(B-L)$ symmetry, for example, then $\nu^{}_j \to \nu^{}_i + \chi$
decays may take place, where $\chi$ is a Goldstone particle (i.e.,
Majoron) \cite{Majoron}. A more exotic scenario, in which massive
neutrinos may decay into unparticles, has also been proposed
\cite{unparticle}.

Here we consider a rather simple case: the decay products of UHE
cosmic neutrinos are invisible, implying that the initial
neutrinos simply disappear. When the neutrino source spectrum
falls with energy in a sufficiently deep way, the daughter
neutrino will also have negligible contributions to the total
neutrino flux. Then the resultant neutrino flavor distribution at
neutrino telescopes is simply given by \cite{decay,Pakvasa1}
\begin{eqnarray}
\phi^{\rm T}_e : \phi^{\rm T}_\mu : \phi^{\rm T}_\tau =
|V^{}_{e1}|^2 : |V^{}_{\mu 1}|^2 : |V^{}_{\tau 1}|^2 \; ,
\end{eqnarray}
provided $\nu^{}_1$ is the lightest neutrino mass eigenstate (and
thus stable). Note that Eq. (18) holds in the assumption that the
heavier neutrinos $\nu^{}_2$ and $\nu^{}_3$ completely decay into
$\nu^{}_1$ and invisible (massless) particles. If the neutrino
mixing matrix $V$ is not unitary, as illustrated in Eq. (7), then
the flavor distribution at neutrino telescopes reads
\begin{eqnarray}
\phi^{\rm T}_e : \phi^{\rm T}_\mu : \phi^{\rm T}_\tau = 4\left(1 -
2W^{}_1\right) : \left(1 - 2W^{}_2 + 4{\rm Re}X\right) : \left(1 -
2W^{}_3 - 4{\rm Re}Y + 2{\rm Re}Z\right) \; .
\end{eqnarray}
It is straightforward to compute the flavor ratios defined in Eq.
(16). In the unitarity limit, we have $R^{}_e = 2$ and $R^{}_\mu =
R^{}_\tau = 1/5$; and in the non-unitary case with only one heavy
Majorana neutrino, we obtain
\begin{eqnarray}
R^{}_e &\approx& 2 - \left[2s^2_{14} - s^2_{24} - s^2_{34} +
4s^{}_{14} s^{}_{24} \cos \varrho - 4 s^{}_{14} s^{}_{34} \cos
\vartheta + 2s^{}_{24} s^{}_{34} \cos (\varrho - \vartheta)\right]
\; , \nonumber
\\
R^{}_\mu &\approx& \frac{1}{5} + \frac{1}{25} \left[4s^2_{14} -
5s^2_{24} + s^2_{34} + 20s^{}_{14} s^{}_{24} \cos \varrho + 4
s^{}_{14} s^{}_{34} \cos \vartheta - 2s^{}_{24} s^{}_{34} \cos
(\varrho - \vartheta)\right] \; , \nonumber
\\
R^{}_\tau &\approx& \frac{1}{5} + \frac{1}{25} \left[4s^2_{14} +
s^2_{24} - 5s^2_{34} - 4s^{}_{14} s^{}_{24} \cos \varrho -
20s^{}_{14} s^{}_{34} \cos \vartheta + 10 s^{}_{24} s^{}_{34} \cos
(\varrho - \vartheta)\right] \; ,
\end{eqnarray}
where $\varrho \equiv \delta^{}_{14} - \delta^{}_{24}$, $\vartheta
\equiv \delta^{}_{14} - \delta^{}_{34}$, and higher-order terms of
$s^{}_{ij}$ have been neglected. The allowed regions of three
flavor ratios are plotted in Figure 2, where the phase angles
$\varrho$ and $\vartheta$ vary freely in the range $[0, 2\pi]$.
Two comments are in order:
\begin{itemize}
\item Different from the case discussed in section 2, here
the deviation of $R^{}_e$ from its value in the unitarity limit
(i.e., $R^{}_e = 2$) can be as large as $4\%$. In comparison, the
deviation of $R^{}_\mu$ or $R^{}_\tau$ from its value in the
unitarity limit (i.e., $R^{}_\mu = R^{}_\tau = 0.2$) can be at the
$0.1\%$ level.

\item It is worth mentioning that additional terms involving
${\rm Re}Y$ and ${\rm Re}Z$ are present in Eq. (19), compared to
Eq. (11). On the other hand, since $s^{}_{14}$ or $s^{}_{24}$ is
confined to a very small value, the non-unitary CP-violating phase
$\varrho$ can hardly affect the flavor ratios in Eq. (17). In the
decay scenario, however, both the phases $\varrho$ and $\vartheta$
can significantly contribute to $R^{}_\alpha$.
\end{itemize}
We see that the flavor distribution of UHE cosmic neutrinos in the
decay scenario is quite different from that in the standard
neutrino oscillation picture. In particular, the democratic flavor
distribution of UHE cosmic neutrinos at neutrino telescopes is
badly broken even if the condition $|V^{}_{\mu i}| = |V^{}_{\tau
i}|$ (for $i=1,2,3$) is satisfied.

\section{Summary}

Assuming that UHE cosmic neutrinos are produced from the decays of
charged pions arising from energetic $pp$ and (or) $p\gamma$
collisions at a distant astrophysical source, one may expect a
democratic flavor distribution $\phi^{\rm T}_e : \phi^{\rm T}_\mu
: \phi^{\rm T}_\tau = 1 : 1 : 1$ at neutrino telescopes if either
$\theta^{}_{13} = 0$ and $\theta^{}_{23} = \pi/4$ (CP invariance)
or $\delta=\pm \pi/2$ and $\theta^{}_{23} = \pi/4$ (CP violation)
are satisfied in the standard parametrization of $V$. A lot of
attention has been focused on small perturbations to the above
conditions such that the resultant flavor distribution is no more
democratic. We have explored a novel possibility, in which $V$ is
non-unitary and its non-unitarity is induced by heavy Majorana
neutrinos as expected in a class of TeV-scale seesaw models, to
examine the flavor distribution of UHE cosmic neutrinos at a
terrestrial neutrino telescope. We have shown that the effect of
unitarity violation on the flavor ratios $\phi^{\rm T}_e :
\phi^{\rm T}_\mu : \phi^{\rm T}_\tau$ can be as large as several
percent. We have also made some brief comments on cosmic neutrino
decays and illustrated the relevant flavor distributions at
neutrino telescopes.

A measurement of the flavor distribution of UHE cosmic neutrinos
is certainly a big challenge to IceCube and other neutrino
telescopes. In the long run, however, we hope that neutrino
telescopes can play an interesting role complementary to the
terrestrial neutrino oscillation experiments in understanding the
intrinsic properties of massive neutrinos and probing possible new
physics.

\vspace{0.25cm}

This work was supported in part by the National Natural Science
Foundation of China.



\begin{figure}[t]
\vspace{-5cm}
\epsfig{file=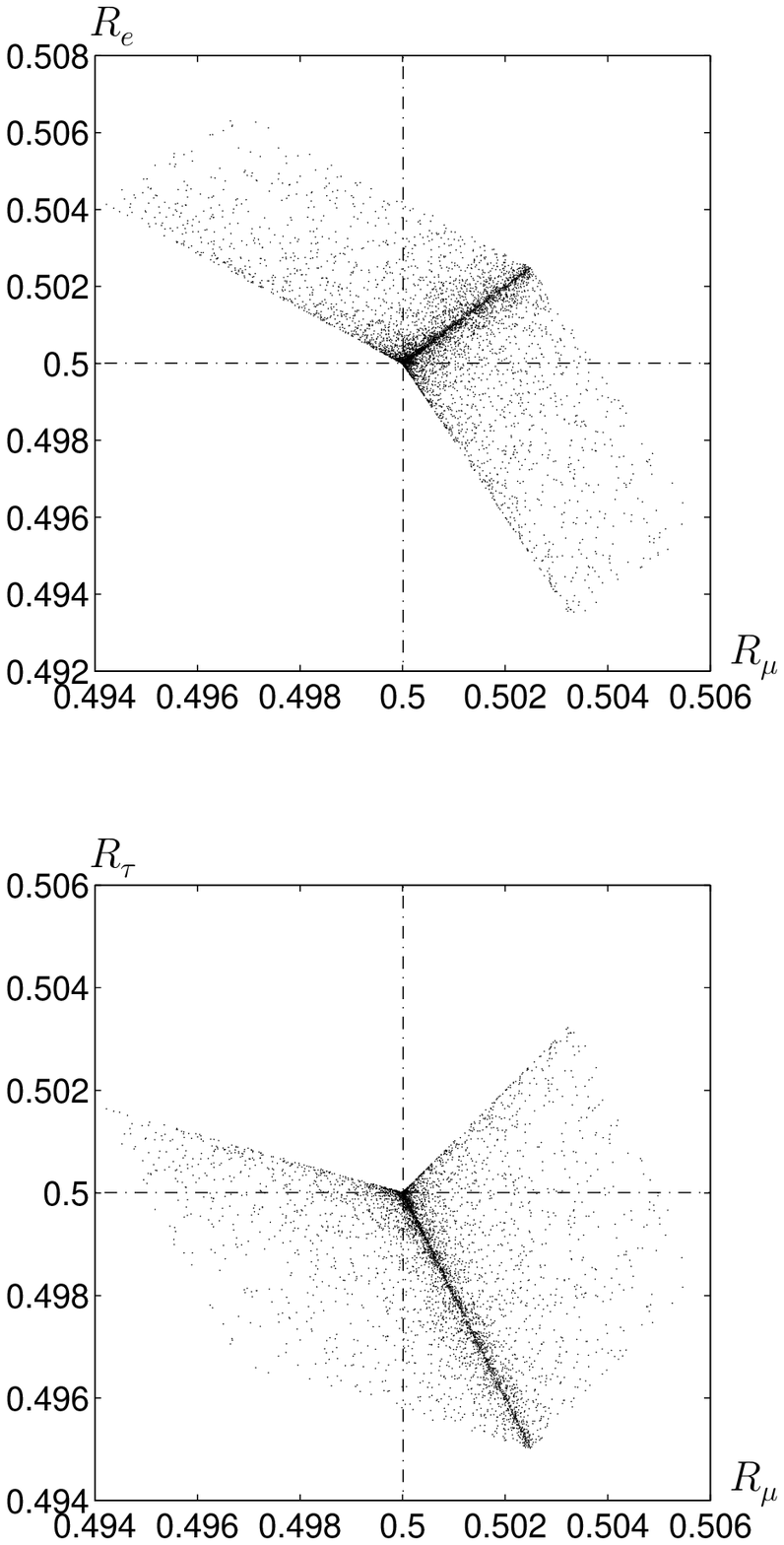,bbllx=0.0cm,bblly=14cm,bburx=10.0cm,bbury=31cm,%
width=10cm,height=17cm,angle=0,clip=0} \vspace{6.5cm}
\caption{Allowed regions of the flavor ratios ($R^{}_e$,
$R^{}_\mu$) and ($R^{}_\tau$, $R^{}_\mu$), where the density of
points is generated by scanning the possible ranges of $s^{}_{ij}$
(for $ij = 14, 24, 34$) according to a flat random number
distribution (i.e., $s^{}_{ij} \in [0, 0.1]$ and
$s^{}_{14}s^{}_{24} < 7.0 \times 10^{-5}$ based on current
experimental constraints on the non-unitarity of $V$).}
\end{figure}

\begin{figure}[t]
\vspace{-5cm}
\epsfig{file=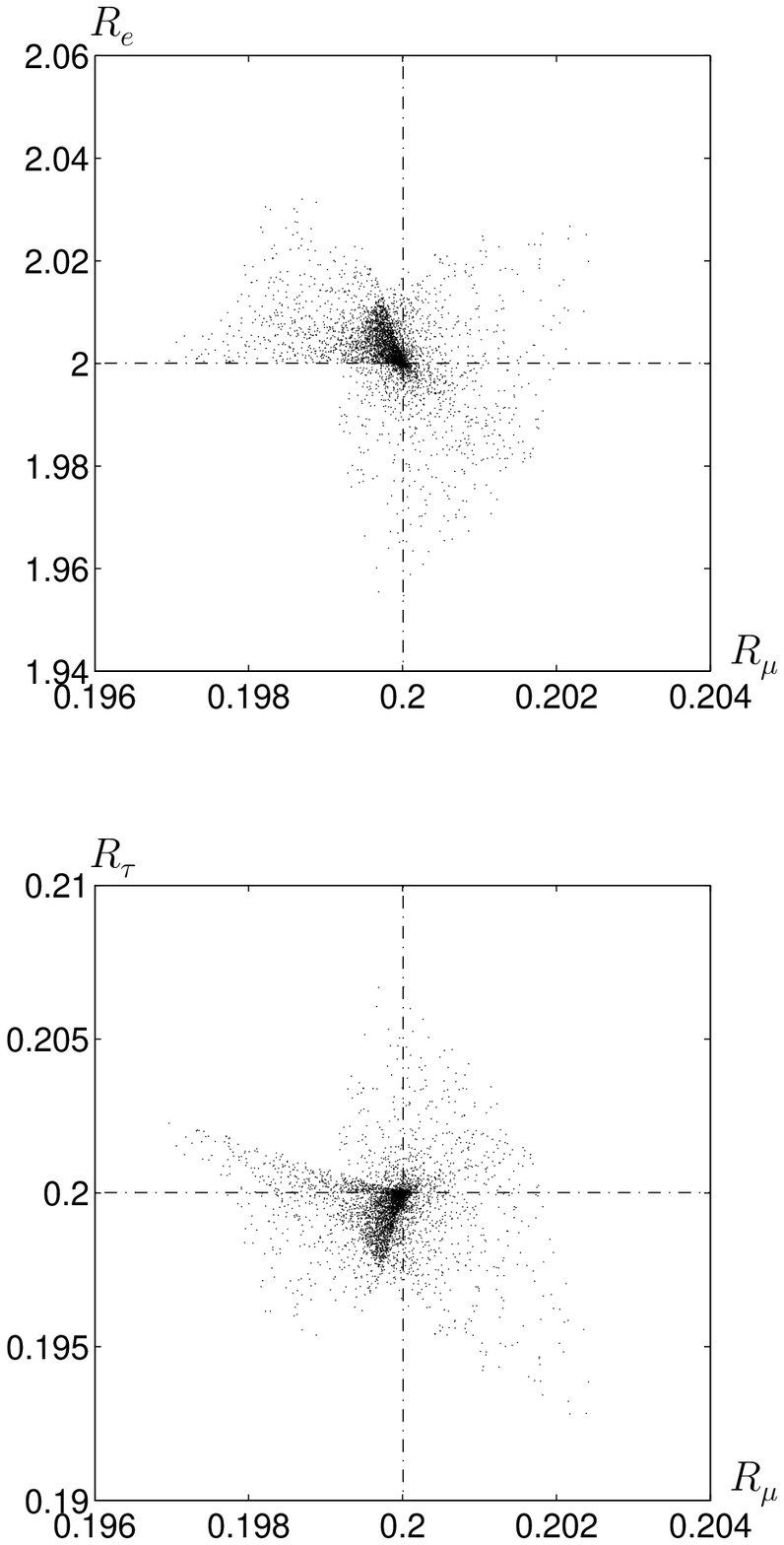,bbllx=0.0cm,bblly=14cm,bburx=10.0cm,bbury=31cm,%
width=10cm,height=17cm,angle=0,clip=0} \vspace{6.5cm}
\caption{Allowed regions of the flavor ratios ($R^{}_e$,
$R^{}_\mu$) and ($R^{}_\tau$, $R^{}_\mu$) in the neutrino decay
scenario, where the density of points is generated by scanning the
possible ranges of $s^{}_{ij}$ (for $ij = 14, 24, 34$) according
to a flat random number distribution (i.e., $s^{}_{ij} \in [0,
0.1]$ and $s^{}_{14}s^{}_{24} < 7.0 \times 10^{-5}$ based on
current experimental constraints on the non-unitarity of $V$).}
\end{figure}


\begin{thebibliography}{99}

\bibitem{SNO} SNO Collaboration, Q.R. Ahmad {\it et al.},
Phys. Rev. Lett. {\bf 89}, 011301 (2002).

\bibitem{SK} For a review, see: C.K. Jung {\it et al.},
Ann. Rev. Nucl. Part. Sci. {\bf 51}, 451 (2001).

\bibitem{KM} KamLAND Collaboration, K. Eguchi {\it et al.},
Phys. Rev. Lett. {\bf 90}, 021802 (2003).

\bibitem{K2K} K2K Collaboration, M.H. Ahn {\it et al.},
Phys. Rev. Lett. {\bf 90}, 041801 (2003).

\bibitem{PDG} Particle Data Group, W.M. Yao {\it et al.},
J. Phys. G {\bf 33}, 1 (2006). See, also, H. Fritzsch and Z.Z.
Xing, Phys. Lett. B {\bf 517}, 363 (2001); Z.Z. Xing, Int. J. Mod.
Phys. A {\bf 19}, 1 (2004).

\bibitem{Vissani} A. Strumia and F. Vissani, hep-ph/0606054.

\bibitem{Symmetry} For recent reviews with extensive references,
see: W. Grimus, hep-ph/0610158; Z.Z. Xing, H. Zhang, and S. Zhou,
Phys. Lett. B {\bf 641}, 189 (2006); T. Baba and M. Yasue,
arXiv:0710.2713 [hep-ph].

\bibitem{IceCube} IceCube Collaboration, J. Ahrens {\it et al.},
Nucl. Phys. Proc. Suppl. {\bf 118}, 388 (2003).

\bibitem{ANTARES} ANTARES Collaboration, E. Aslanides {\it et al.},
astro-ph/9907432.

\bibitem{NESTOR} NESTOR Collaboration, S.E. Tzamarias {\it et al.},
Nucl. Instrum. Meth. A {\bf 502}, 150 (2003).

\bibitem{NEMO} NEMO Collaboration, P. Piatelli, Nucl. Phys. Proc. Suppl.
{\bf 143}, 359 (2005).

\bibitem{Pakvasa} J.G. Learned and S. Pakvasa, Astropart. Phys.
{\bf 3}, 267 (1995).

\bibitem{Serpico} H. Athar, M. Jezabek, and O. Yasuda, Phys. Rev.
D {\bf 62}, 103007 (2000); L. Bento, P. Keranen, and J. Maalampi,
Phys. Lett. B {\bf 476}, 205 (2000); G.J. Gounaris and G. Moultaka,
hep-ph/0212110; Y. Farzan and A.Yu. Smirnov, Phys. Rev. D {\bf 65},
113001 (2002); P. Keranen, J. Maalampi, M. Myyrylainen, and J.
Riittinen, Phys. Lett. B {\bf 574}, 162 (2003); P.D. Serpico and M.
Kachelrie$\rm \ss$, Phys. Rev. Lett. {\bf 94}, 211102 (2005); P.
Bhattacharjee and N. Gupta, hep-ph/0501191; P.D. Serpico, Phys. Rev.
D {\bf 73}, 047301 (2006).

\bibitem{Xing06} Z.Z. Xing, Phys. Rev. D {\bf 74}, 013009 (2006);
Z.Z. Xing and S. Zhou, Phys. Rev. D {\bf 74}, 013010 (2006).

\bibitem{NT} W. Winter, Phys. Rev. D {\bf 74}, 033015 (2006); H. Athar,
C.S. Kim, and J. Lee, Mod. Phys. Lett. A {\bf 21}, 1049 (2006); W.
Rodejohann, JCAP {\bf 0701}, 029 (2007); Z.Z. Xing, Nucl. Phys. B
(Proc. Suppl.) {\bf 168}, 274 (2007); K. Blum, Y. Nir, and E.
Waxman, arXiv:0706.2070 [hep-ph]; P. Lipari, M. Lusignoli, and D.
Meloni, Phys. Rev. D {\bf 75}, 123005 (2007); D. Majumdar and A.
Ghosal, Phys. Rev. D {\bf 75}, 113004 (2007); R.L. Awasthi and S.
Choubey, Phys. Rev. D {\bf 76}, 113002 (2007); G.R. Hwang and S.
Kim, arXiv:0711.3122 [hep-ph]; Z.Z. Xing, Nucl. Phys. B (Proc.
Suppl.) {\bf 175-176}, 421 (2008); S. Pakvasa, W. Rodejohann, and
T.J. Weiler, JHEP {\bf 0802}, 005 (2008); S. Choubey, V. Niro, and
W. Rodejohann, arXiv:0803.0423 [hep-ph].

\bibitem{decay} J.F. Beacom, N.F. Bell, D. Hooper, S. Pakvasa, and
T.J. Weiler, Phys. Rev. Lett. {\bf 90}, 181301 (2003); G.
Barenboim and C. Quigg, Phys. Rev. D {\bf 67}, 073024 (2003); J.F.
Beacom, N.F. Bell, D. Hooper, S. Pakvasa, and T.J. Weiler, Phys.
Rev. D {\bf 68}, 093005 (2003); Phys. Rev. D {\bf 69}, 017303
(2004); D. Meloni and T. Ohlsson, Phys. Rev. D {\bf 75}, 125017
(2007); S. Pakvasa, arXiv:0803.1701 [hep-ph]; M. Maltoni and W.
Winter, arXiv:0803.2050 [hep-ph].

\bibitem{XZ08} Z.Z. Xing and S. Zhou, arXiv:0804.3512v2 [hep-ph] (unpublished).

\bibitem{TB} P.F. Harrison, D.H. Perkins, and W.G. Scott, Phys.
Lett. B {\bf 530}, 167 (2002); Z.Z. Xing, Phys. Lett. B {\bf 533},
85 (2002); P.F. Harrison and W.G. Scott, Phys. Lett. B {\bf 535},
163 (2002); X.G. He and A. Zee, Phys. Lett. B {\bf 560}, 87 (2003).

\bibitem{TT} Z.Z. Xing, arXiv:0805.0416 [hep-ph], to appear in Phys. Rev. D
(Rapid Communication).

\bibitem{SS1} P. Minkowski, Phys. Lett. B {\bf 67}, 421 (1977);
T. Yanagida, in {\it Proceedings of the Workshop on Unified Theory
and the Baryon Number of the Universe}, edited by O. Sawada and A.
Sugamoto (KEK, Tsukuba, 1979); M. Gell-Mann, P. Ramond, and R.
Slansky, in {\it Supergravity}, edited by P. van Nieuwenhuizen and
D. Freedman (North Holland, Amsterdam, 1979); S. L. Glashow, in {\it
Quarks and Leptons}, edited by M. L$\acute{\rm e}$vy {\it et al.}
(Plenum, New York, 1980); R. N. Mohapatra and G. Senjanovic, Phys.
Rev. Lett. {\bf 44}, 912 (1980).

\bibitem{SS2} M. Magg and C. Wetterich, Phys. Lett. B {\bf 94}, 61
(1980); J. Schechter and J.W.F. Valle, Phys. Rev. D {\bf 22}, 2227
(1980); T.P. Cheng and L.F. Li, Phys. Rev. D {\bf 22}, 2860 (1980);
R.N. Mohapatra and G. Senjanovic, Phys. Rev. D {\bf 23}, 165 (1981).

\bibitem{XZ} See, e.g., Z.Z. Xing and S. Zhou, High Energy Phys. Nucl. Phys.
{\bf 30}, 828 (2006); W. Chao, S. Luo, Z.Z. Xing, and S. Zhou,
Phys. Rev. D {\bf 77}, 016001 (2008); W. Chao, Z. Si, Z.Z. Xing,
and S. Zhou, arXiv:0804.1265 [hep-ph]; P. Ren and Z.Z. Xing,
arXiv:0805.4292 [hep-ph]; W. Chao, arXiv:0806.0889 [hep-ph].

\bibitem{Antusch} S. Antusch, C. Biggio, E. Fernandez-Martinez,
M.B. Gavela, and J. Lopez-Pavon, JHEP {\bf 0610}, 084 (2006).

\bibitem{Xing08} Z.Z. Xing, Phys. Lett. B {\bf 660}, 515 (2008).

\bibitem{Yasuda} E. Fernandez-Martinez, M.B. Gavela,
J. Lopez-Pavon, and O. Yasuda, Phys. Lett. B {\bf 649}, 427
(2007); J. Lopez-Pavon, AIP Conf. Proc. {\bf 981}, 219 (2008); S.
Goswami and T. Ota, arXiv:0802.1434 [hep-ph]; S. Luo,
arXiv:0804.4897 [hep-ph].

\bibitem{Zralek} M. Czakon, J. Gluza, and M. Zralek, Acta Phys. Polon. B {\bf 32},
3735 (2001). B. Bekman, J. Gluza, J. Holeczek, J. Syska, and M.
Zralek, Phys. Rev. D {\bf 66}, 093004 (2002); J. Holeczek, J.
Kisiel, J. Syska, and M. Zralek, Eur. Phys. J. C {\bf 52}, 905
(2007).

\bibitem{Sterile} H. Athar, M. Jezabek, and O. Yasuda, in Ref. \cite{Serpico};
R.L. Awasthi and S. Choubey, in Ref. \cite{NT}.

\bibitem{Mtype2} P.H. Gu, H. Zhang, and S. Zhou, Phys. Rev. D {\bf
74}, 076002 (2006); A.H. Chan, H. Fritzsch, S. Luo, and Z.Z. Xing,
Phys. Rev. D {\bf 76}, 073009 (2007).

\bibitem{Majoron} Y. Chikashige, R.N. Mohapatra, and R.D. Peccei, Phys.
Lett. B {\bf 98}, 265 (1981); G.B. Gelmini and M. Roncadelli, Phys.
Lett. B {\bf 99}, 411 (1981); V.D. Barger, W.Y. Keung, and S.
Pakvasa, Phys. Rev. D {\bf 25}, 907 (1982); J.W.F. Valle, Phys.
Lett. B {\bf 131}, 87 (1983).

\bibitem{unparticle} S. Zhou, Phys. Lett. B {\bf 659}, 336 (2008);
S.L. Chen, X.G. He, and H.C. Tsai, JHEP {\bf 0711}, 010 (2007);
X.Q. Li, Y. Liu, and Z.T. Wei, arXiv:0707.2285 [hep-ph]; D.
Majumdar, arXiv:0708.3485 [hep-ph].

\bibitem{Pakvasa1} S. Pakvasa, hep-ph/0305317; and references
therein.
\end{thebibliography}
\end{document}